\documentclass[preprint,showpacs,preprintnumbers,amsmath,amssymb]{revtex4}
\usepackage{graphicx}
\usepackage{dcolumn}
\usepackage{bm}

\def\lesssim{\ \raise.3ex\hbox{$<$}\kern-0.8em\lower.7ex\hbox{$\sim$}\ }
\def\gesim{\ \raise.3ex\hbox{$>$}\kern-0.8em\lower.7ex\hbox{$\sim$}\ }
\font\scripti=cmmi7
\font\scriptscripti=cmmi5
\def\sib#1{\setbox0 = \hbox{\scripti #1}
  \kern-.02em\copy0\kern-\wd0
  \kern.04em\box0} 
\def\ssib#1{\setbox0 = \hbox{\scriptscripti #1}
  \kern-.02em\copy0\kern-\wd0
  \kern.04em\box0} 
\font\tenib=cmmib10 
\skewchar\tenib='177 \skewchar\tenib='177 \skewchar\tenib='177
\def\pbold#1{\setbox0 = \hbox{$ #1 $}
  \kern-.022em\copy0\kern-\wd0
  \kern.011em\copy0\kern-\wd0
  \kern.011em\copy0\kern-\wd0
  \kern.011em\copy0\kern-\wd0
  \kern.011em\box0} 
\begin{document}
\title{Itinerant-localized dual character of a strongly-correlated superfluid Bose gas in an optical lattice}
\author{Y. Ohashi, M. Kitaura, and H. Matsumoto}
\affiliation{Institute of Physics, University of Tsukuba, Tsukuba, Ibaraki 305, Japan}
\date{\today}
\begin{abstract}
We investigate a strongly-correlated Bose gas in an optical lattice. Extending the standard-basis operator method developed by Haley and Erd\"os to a boson Hubbard model, we calculate excitation spectra in the superfluid phase, as well as in the Mott insulating phase, at $T=0$. In the Mott phase, the excitation spectrum has a finite energy gap, reflecting the localized character of atoms. In the superfluid phase, the excitation spectrum is shown to have an itinerant-localized dual structure, where the gapless Bogoliubov mode (which describes the itinerant character of superfluid atoms) and a band with a finite energy gap coexist. We also show that the rf-tunneling current measurement would give a useful information about the duality of a strongly-correlated superfluid Bose gas near the superfluid-insulator transition. 
\end{abstract}
\maketitle
%
\par
Recently, a strongly-correlated Bose has been realized in an optical lattice\cite{Greiner,Esslinger}. When the optical lattice potential is strong enough, the system can be described by a boson Hubbard model\cite{Fisher,Sachdev}, where atoms are hopping between nearest-neighbor lattice sites, interacting with each other with the on-site repulsion $U$. In a cold atom gas, the hopping matrix element $t$ is tunable by adjusting the strength of laser light, so that one can realized the strongly-correlated regime characterized by $U/zt \gesim 1$ (where $z$ is the number of nearest neighbor sites). Theoretical work on the boson Hubbard model\cite{Fisher,Kotlier,Krauth,Krish,Monien,Stoof,Stoof2,Burnett,Dupuis} has predicted the superfluid-Mott insulator (SI) transition, which has been recently confirmed in $^{87}$Rb Bose gas\cite{Greiner,Esslinger}. The experimental result shows that the SI transition occurs at $U/zt\sim 6$ in a cubic lattice, which is in good agreement with the theoretical prediction $U/zt=3+2\sqrt{2}=5.83$\cite{Krish,Stoof}. More recently, the optical lattice has been also applied to a cold Fermi gas\cite{Kohl}, and the Brillouin zone has been observed.
\par
In the superfluid phase near the SI transition, while superfluid atoms are itinerant, we can expect that the system would also have an insulating (localized) character. Indeed, an itinerant-localized dual character has been obtained in a {\it fermion} Hubbard model\cite{Bulla}, where an {\it itinerant} heavy fermion band at the Fermi level and upper- and lower-Hubbard bands, describing the {\it localized} character of fermions, coexist. At the metal-insulator transition, the itinerant heavy fermion band vanishes. In the Mott insulating phase, the upper and lower Hubbard bands only remains, that are separated by the so-called Hubbard gap $E_g\sim U$. Thus, it is a very interesting problem how this kind of duality can be seen in a strongly-correlated superfluid {\it Bose} gas. For this purpose, cold atom gases in optical lattices are very useful, because various physical parameters, such as the number density $n$ of atoms and the interaction $U/zt$, are tunable. Since the SI transition is absent in a weakly-interacting Bose gas, the duality is a characteristic phenomenon where the strong interaction between atoms plays an essential role. 
\par

\begin{figure}
\includegraphics[width=6.5cm,height=4.5cm]{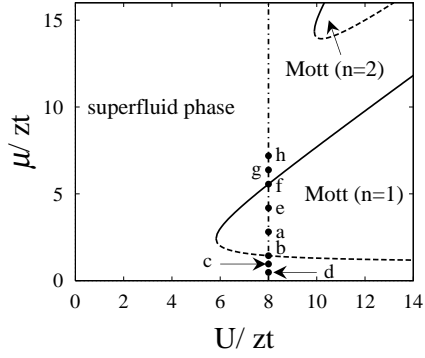}%
\caption{
Phase diagram of the boson Hubbard model. The SI phase boundary shown by the solid (dashed) line is given by ${\bar \mu}_+$ (${\bar \mu}_-$) in (\ref{eq.11}). The solid circles at $U/zt=8$ (a$\sim$h) are the positions where excitation spectrum and the dispersion relation are calculated in Figs. 2-4. $\mu/zt$ at each position equals
(a) ${\bar \mu}_-+({\bar \mu}_+-{\bar \mu}_-)/3$ $(n=1)$,
(b) ${\bar \mu}_-$ $(n=1)$,
(c) $2{\bar \mu}_-/3$ $(n=0.89)$, 
(d) ${\bar \mu}_-/3$ $(n=0.74)$,
(e) ${\bar \mu}_-+2({\bar \mu}_+-{\bar \mu}_-)/3$ $(n=1)$,
(f) ${\bar \mu}_+$ $(n=1)$,
(g) ${\bar \mu}_++(U-{\bar \mu}_+)/3$ $(n=1.16)$, and  
(h) ${\bar \mu}_++2(U-{\bar \mu}_+)/3$ $(n=1.36)$. 
\label{fig1} 
}
\end{figure}

In this paper, we discuss excitation properties of a strongly-correlated Bose gas near the SI transition. Extending the standard basis operator method developed by Haley and Erd\"os\cite{Haley} to the boson Hubbard model, we calculate the superfluid density of states at $T=0$. This method has been recently used by Sheshadri\cite{Krish}, Kanobe\cite{Kanobe} and their coworkers to study the SI transition and the rf-spectrum in the Mott phase, respectively. In the superfluid phase near the SI transition, we show the coexistence of the itinerant gapless Bogoliubov mode and a band with a finite threshold energy, which reflects the localized character of atomic states. We also show that this novel phenomenon may be observed by the rf-tunneling measurement. 
\par
We consider a superfluid Bose gas in a cubic optical lattice, described by the boson Hubbard model
\begin{eqnarray}
H=-t\sum_{(i,j)}c^\dagger_ic_j-{U \over 2}\sum_i {\hat n}_i({\hat n}_i-1)-
\mu\sum_i {\hat n}_i.
\label{eq.1}
\end{eqnarray}
Here, $c_i^\dagger$ is a creation operator of a Bose atom at the $i$-th site, and ${\hat n}_i\equiv c_i^\dagger c_i$. In the first term, $t$ describes the nearest-neighbor hopping, where the summation $(i,j)$ is taken over the nearest-neighbor sites. $U$ is the on-site repulsion, and $\mu$ is the chemical potential. For simplicity, we ignore effects of a trap potential in (\ref{eq.1}). 
\par
In the superfluid phase, we divide the Bose operator $c_i$ into the condensate $\Phi\equiv \langle c_i\rangle$ and the non-condensate $\delta c_i\equiv c_i-\Phi$. (We take $\Phi$ to be real.) Substituting the expression $c_i=\Phi+\delta c_i$ into (\ref{eq.1}), we obtain $H=\sum_i h_i-t\sum_{(i,j)}\delta c^\dagger_i\delta c_j$, where $h_i=zt\Phi^2-zt\Phi(c_i+c_i^\dagger)+{U \over 2}{\hat n}_i({\hat n}_i-1)-\mu{\hat n}_i$ (where $z=6$ is the number of nearest-neighbor sites of the cubic lattice). The BEC order parameter $\Phi$ is determined by minimizing the ground state energy $E_{\alpha=0}$ of the on-site Hamiltonian $h_i$\cite{Krish}. Using the eigenstates $|i,\alpha\rangle$ of $h_i$ (with eigenenergies $E_\alpha$), we rewrite the Hamiltonian as
\begin{eqnarray}
H=\sum_{i,\alpha}E_\alpha {\hat L}^i_{\alpha,\alpha'}
-t\sum_{(i,j) \atop \alpha,\alpha',\beta,\beta'}T^{i,j}_{\alpha,\alpha',\beta,\beta'}
\delta{\hat L}^i_{\alpha,\alpha'}
\delta{\hat L}^j_{\beta,\beta'},
\label{eq.2}
\end{eqnarray}
where $T^{i,j}_{\alpha,\alpha',\beta,\beta'}\equiv \langle i,\alpha|c^\dagger_i|i,\alpha'\rangle\langle j,\beta|c_j|j,\beta'\rangle$. ${\hat L}^i_{\alpha\alpha'}\equiv |i,\alpha\rangle\langle i,\alpha'|$ is referred to as the standard basis operators\cite{Haley}, and $\delta{\hat L}^i_{\alpha,\alpha'}\equiv {\hat L}^i_{\alpha,\alpha'}-\langle {\hat L}^i_{\alpha,\alpha'}\rangle$, where $\langle\cdot\cdot\cdot\rangle$ is the expectation value in the ground state $|i,\alpha=0\rangle$. 
\par
To calculate the excitation spectrum, it is convenient to introduce the single-particle Green's function, defined by $g_{i,j}(\tau)=-i\Theta(\tau)\langle[c_i(\tau),c^\dagger_j]\rangle$ [where $\Theta(\tau)$ is the step function]. In the ${\hat L}^i_{\alpha,\alpha'}$-representation, we obtain $g_{i,j}(\tau)=\sum_{\alpha,\alpha',\beta,\beta'}T^{j,i}_{\beta,\beta',\alpha,\alpha'}G^{i,j}_{\alpha,\alpha',\beta,\beta'}(\tau)$, where 
\begin{eqnarray}
G^{i,j}_{\alpha,\alpha',\beta,\beta'}(\tau)=
-i\Theta(\tau)\langle[{\hat L}^i_{\alpha,\alpha'}(\tau),{\hat L}^j_{\beta,\beta'}]
\rangle.
\label{eq.4}
\end{eqnarray}
In the random phase approximation (RPA) in terms of the hopping $t$, the equation of motion for $G$ is given by\cite{Haley,Krish}, in the energy ($\omega$) and momentum (${\bf p}$) space,
\begin{eqnarray}
\delta_{\alpha,\beta'}\delta_{\alpha',\beta}D_{\alpha,\alpha'}
&=&
-\varepsilon_{\bf p}D_{\alpha,\alpha'}
\sum_{\gamma,\gamma'}
{\tilde T}_{\alpha',\alpha,\gamma,\gamma'}G_{\gamma,\gamma',\beta,\beta'}({\bf p},\omega)\nonumber
\\
&+&
[\omega-(E_{\alpha'}-E_\alpha)]G_{\alpha,\alpha',\beta,\beta'}({\bf p},\omega).
\label{eq.5}
\end{eqnarray}
Here, $\varepsilon_{\bf p}\equiv -2t\sum_{j=x,y,z}\cos(p_j)$ describes the tunneling effect between sites (where the lattice constant is taken to be unity). $D_{\alpha\alpha'}\equiv\langle L_{\alpha,\alpha}\rangle-\langle L_{\alpha',\alpha'}\rangle$, and ${\tilde T}_{\alpha',\alpha,\gamma,\gamma'}\equiv T^{i,j}_{\alpha',\alpha,\gamma,\gamma'}+T^{j,i}_{\gamma,\gamma',\alpha',\alpha}$, where $(i,j)$ is a nearest-neighbor pair. Substituting the solution of (\ref{eq.5}) into the single-particle Green's function $g({\bf p},\omega)$ in the energy and momentum space, we obtain
\begin{eqnarray}
g({\bf p},\omega)=
{\Pi({\bf p},\omega) \over 1-\varepsilon_{\bf p}\Pi({\bf p},\omega)}.
\label{eq.6}
\end{eqnarray}
Here,
\begin{eqnarray}
\Pi({\bf p},\omega)=A_{11}(\omega)+\varepsilon_{\bf p}
{A_{12}(\omega)A_{21}(\omega) \over 1-\varepsilon_{\bf p}A_{22}(\omega)},
\label{eq,7}
\end{eqnarray}
\begin{eqnarray}
\left(
\begin{array}{cc}
A_{11}&A_{12}\\
A_{21}&A_{22}
\end{array}
\right)
&=&
\sum_\alpha
{
\left(
\begin{array}{cc}
y_{0,\alpha}y^*_{\alpha,0}&
y_{0,\alpha}y_{\alpha,0}\\
y^*_{0,\alpha}y^*_{\alpha,0}&
y^*_{0,\alpha}y_{\alpha,0}
\end{array}
\right)
\over \omega_++(E_0-E_\alpha)}
\nonumber
\\
&-&
\sum_\alpha
{ 
\left(
\begin{array}{cc}
y_{\alpha,0}y^*_{0,\alpha}&
y_{\alpha,0}y_{0,\alpha}\\
y^*_{\alpha,0}y^*_{0,\alpha}&
y^*_{\alpha,0}y_{0,\alpha}
\end{array}
\right)
\over \omega_++(E_\alpha-E_0)},
\label{eq.8}
\end{eqnarray}
where $\omega_+=\omega+i\delta$. We note that $y_{\alpha,\beta}\equiv\langle i,\alpha|c_i|i,\beta\rangle$ is independent of the position $i$.
\par
\begin{figure}
\includegraphics[width=8cm,height=7cm]{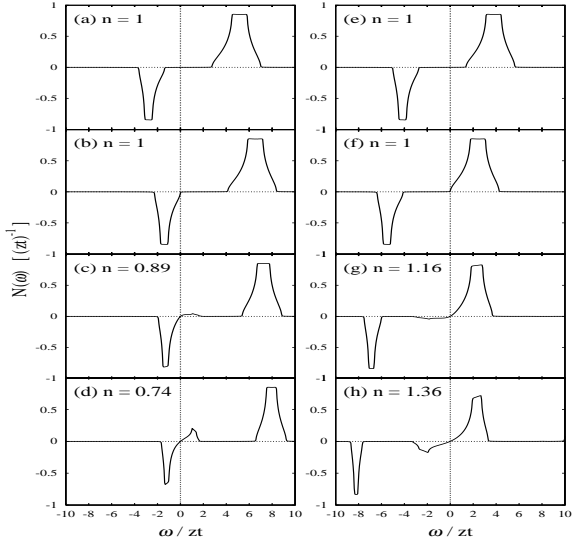}%
\caption{
Excitation spectrum $N(\omega)$ near the SI phase boundary at $U/zt=8$. The value of $\mu$ in each panel is shown in Fig. 1. (a) and (e): Mott phase. (b) and (f): SI phase transition. (c) and (d): Superfluid phase with $n<1$. (g) and (h): Superfluid phase with $n>1$. We have introduced a small imaginary part $\Gamma/zt=0.001$ to the energies, to smooth out the results.
\label{fig2} 
}
\end{figure}

In the Mott phase ($\Phi=0$), the number density of atoms $n$ per site is an integer, and the eigenstates $|i,\alpha\rangle$ of $h_i$ are simply given by the number states, $|i,n\rangle={1 \over \sqrt{n!}}(c_i^\dagger)^n|{\rm vacuum}\rangle$ with $E_n=-\mu n+{U \over 2}n(n-1)$. Among them, $|i,n=n_0\rangle$ is the ground state when $U(n_0-1)\le\mu<Un_0$.  Noting that $y_{n,n'}=\sqrt{n'}\delta_{n,n'-1}$, we obtain the Green's function in the Mott phase as
\begin{eqnarray}
g({\bf p},\omega)={A_{11}(\omega) \over 1-\varepsilon_{\bf p}A_{11}(\omega)},
\label{eq.9}
\end{eqnarray}
where
\begin{eqnarray}
A_{11}=
{n+1 \over \omega_+-\Delta E_{n+1,n}}-
{n \over \omega_++\Delta E_{n-1,n}}.
\label{eq.10}
\end{eqnarray}
Here, $\Delta E_{n+1,n}\equiv E_{n+1}-E_n=Un-\mu$ is the excitation energy by adding an atom to the ground state when $t=0$, while $\Delta E_{n-1,n}\equiv E_{n-1}-E_n=\mu-U(n-1)$ is the energy to extract an atom from the ground state. $\Delta E_{n-1,n}$ may be regarded as the energy to add a {\it hole} to the ground state. The SI transition is determined by the condition when (\ref{eq.9}) has a gapless excitation at ${\bf p}=0$, which gives\cite{Fisher,Krish,Stoof,Kanobe} 
\begin{eqnarray}
{\mu \over zt}&=&{1 \over 2}
\Bigl[(2n-1){U \over zt}-1\pm
\sqrt{\Bigl({U \over zt}\Bigr)^2-2(2n+1){U \over zt}+1}
\Bigr]
\nonumber
\\
&\equiv&
{\bar \mu}_\pm.
\label{eq.11}
\end{eqnarray}
The lowest value of $U/zt$ to realize the Mott phase equals $U/zt=3+2\sqrt{2}=5.83$ $(n=1)$\cite{Greiner,Krish,Stoof}. We show the SI phase boundary given by (\ref{eq.11}) in Fig. 1. 
\par
\begin{figure}
\includegraphics[width=6.5cm,height=4.5cm]{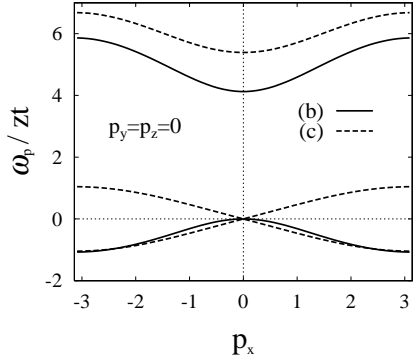}%
\caption{
Dispersion relation in the $p_x$ direction. (b) SI phase boundary. (c) Superfluid phase. 
\label{fig3} 
}
\end{figure}

\par
Now we consider excitation properties near the SI phase boundary. Figure 2 shows the density of states defined by
\begin{eqnarray}
N(\omega)=-{1 \over \pi}\sum_{\bf p}{\rm Im}[g({\bf p},\omega)].
\label{eq.12}
\end{eqnarray}
In Fig. 2, the positive energy region describes {\it particle-type} excitations, and {\it hole-type} excitations are given by the spectrum in the negative energy region. We note that $N(\omega)<0$ for $\omega<0$. [Since the Bose distribution function $n_B(\omega)$ is negative for $\omega<0$, the number of particles $N(\omega)n_B(\omega)$ at $\omega$ is always positive.] In the Mott phase with $n=1$ [panels (a) and (e)], a finite energy gap exists in particle and hole excitations. When $t=0$, the energy gap $E_g$ between particle and hole excitations equals $E_g=\Delta E_{10}+\Delta E_{01}=U$, so that $E_g$ is found to correspond to the {\it Hubbard gap} in the fermion Hubbard model. Namely, the presence of this gap reflects the localized character of Bose atoms in the optical lattice. The finite widths of the particle and hole bands in Fig. 2 are due to the nearest neighbor hopping $t$. At the lower SI phase boundary [dashed line in Fig. 1], Fig. 2(b) shows that the hole excitations become gapless, which means that holes become itinerant. On the other hand, excitations in the particle branch ($\omega>0$) still has a gap in Fig. 2(b). However, as one enters the superfluid phase [panels (c) and (d)], a gapless band suddenly appears in the particle branch. The resulting excitation spectrum in the particle branch has a {\it dual} structure, where the gapless excitation spectrum and the band with a finite threshold energy coexist. We note that this phenomenon can be also seen when one enters the superfluid phase through the upper SI phase boundary [solid line in Fig. 1]. In this case, as shown in Figs. 2(e)-(h), particle excitations become gapless at the SI phase transition, and the dual band structure is obtained in the hole branch. 
\par
Figure 3 shows the dispersion relation near the lower SI phase boundary. Although the spectral weights of the gapless particle and hole excitations are different in Fig. 2(c), their dispersion relations are {\it symmetric} with respect to $\omega=0$. Their linear dispersions around $p_x=0$ indicate that they are the Bogoliubov modes. At the SI transition, the gapless dispersion in the particle branch vanishes, and the hole dispersion becomes quadratic, reflecting the disappearance of superfluidity.
\par
The reason for the sudden appearance of the gapless spectrum in the superfluid phase is that the macroscopic wave function of the superfluid phase is a coherent state, involving the various number states $|n\rangle$ ($n=0,1,2,\cdot\cdot\cdot$). This leads to the hybridization of the particle and hole excitations in the superfluid phase. As a result, the Bogoliubov excitation from the condensate is accompanied by both particle and hole excitations, which leads to the gapless particle and hole dispersions shown in Fig. 3 (dashed lines). 
\par
We note that the gapless excitations in the superfluid phase can be studied analytically in the strongly-correlated limit ($U/zt\gg 1$). In this extreme case, when $n<1$, we may only take into account the two number states, $|n=0\rangle$ and $|n=1\rangle$, because of the strong repulsion. The resulting single-particle Green's function has the simple form
\begin{eqnarray}
g({\bf p},\omega)
&=&
{1 \over 2}
{
{(1-2n_{\rm c})zt+(1-4n_{\rm c})\varepsilon_{\bf p} \over \Omega_{\bf p}}-(2n-1)
\over \omega_+-\Omega_{\bf p}
}
\nonumber
\\
&+&
{1 \over 2}
{
{(1-2n_{\rm c})zt+(1-4n_{\rm c})\varepsilon_{\bf p} \over \Omega_{\bf p}}+(2n+1)
\over \omega_++\Omega_{\bf p}
},
\label{eq.13}
\end{eqnarray}
where $n_{\rm c}\equiv\Phi^2=n(1-n)$ is the condensate fraction. (\ref{eq.13}) shows the existence of the gapless dispersions in the particle and hole branches, given by $\omega=\pm\Omega_{\bf p}=\pm\sqrt{(zt+(2n-1)^2\varepsilon_{\bf p})(zt+\varepsilon_{\bf p})}$. In the long-wave length limit, we obtain $\omega=\pm v_\phi p$. Here, the sound velocity is given by $v_\phi/zt=\Phi\sqrt{2/3}=\sqrt{2n_{\rm c}/3}$\cite{note}, which is independent of the interaction $U$. This result is quite different from the usual Bogoliubov sound velocity in a weakly-interacting superfluid Bose gas, given by $v_\phi=\sqrt{Un_{\rm c}/M}\propto \sqrt{U}$ (where $M$ is the mass of an atom). We note that the spectral weight of the particle branch ($\omega>0$) in (\ref{eq.13}) vanishes at the SI phase boundary ($n_{\rm c}=0$ and $n=1$), as expected from Fig. 2(b). 
\par
To observe excitation properties of a superfluid Bose gas, the rf-tunneling current spectroscopy is a powerful tool\cite{Luxat}. In this measurement, atoms are transferred into another hyperfine state (described by a Bose operator $b_{\bf p}$) by laser radiation. In the rotational wave approximation\cite{Luxat}, this process is described by the tunneling Hamiltonian
$
H_T=\gamma
\sum_{\bf p}
\Bigl[
e^{-i\omega_Lt}b^\dagger_{{\bf p}+{\bf q}_L}c_{\bf p}+h.c.
\Bigr]
$, 
where ${\bf q}_L$ and $\omega_L$ are the momentum and frequency of the laser light, respectively. $\gamma$ is the tunneling matrix element between the two hyperfine states described by $b_{\bf p}$ and $c_{\bf p}$. The Hamiltonian of the $b$-state has the form $H_b=\sum_{\bf p}[\varepsilon_{\bf p}^b+\omega_b-\mu_b]b_{\bf p}^\dagger b_{\bf p}$, where $\varepsilon_{\bf p}^b=-2t\sum_{j=x,y,z}\cos p_j+zt$. $\omega_b$ and $\mu_b$ represent the threshold energy of the $b$-band and the chemical potential, respectively. In this Hamiltonian, we have ignored the interaction in the $b$-state and have assumed that the hopping matrix element $t$ is the same as that of the $c$-state, for simplicity. In the first order perturbation in terms of $H_T$, we obtain the rf-tunneling current $I(\omega)$ from the $c$-state to the $b$-state as (we set ${\bf q}_L=0$)
\begin{eqnarray}
I(\omega)=I_{\rm c}\delta(\omega)+
2\gamma^2\sum_{\bf p}
{\rm Im}[g({\bf p},\varepsilon_{\bf p}^b-\omega)]\Theta(\omega-\varepsilon_{\bf p}^b).
\label{eq.15}
\end{eqnarray}
Here, $\omega\equiv \omega_L-\omega_b+\mu_b$ is a renormalized detuning energy, and $I_{\rm c}\equiv 2\pi\gamma^2n_{\rm c}$ describes the contribution from the condensate. The second term in (\ref{eq.15}) involves information about the excitation spectrum shown in Fig. 2. Because of the step function in (\ref{eq.15}), the second term is dominated by hole excitations at $T=0$\cite{Kanobe}. 
\par

\begin{figure}
\includegraphics[width=8cm,height=7cm]{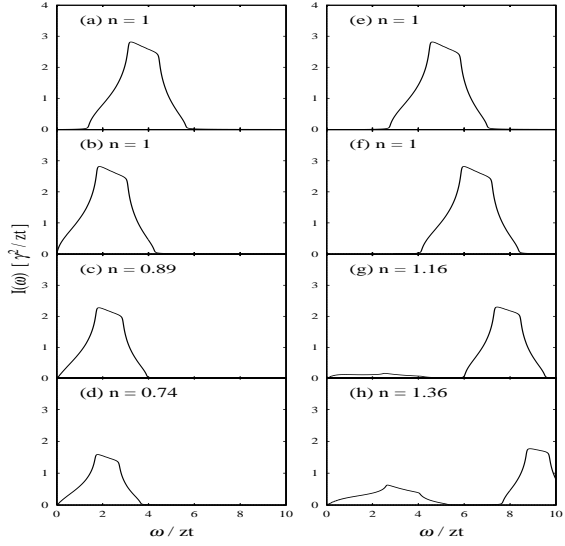}%
\caption{
Low energy rf-spectrum at $T=0$. The contribution from the condensate $I_{\rm c}$ at $\omega=0$ is not shown in this figure. We have introduced a small imaginary part $\Gamma/zt=0.03$ to the energies, to smooth out the results.
\label{fig4} 
}
\end{figure}

Figure 4 shows the rf-spectrum near the Mott phase with $n=1$. Near the lower SI boundary [dashed line in Fig. 1], the excitation gap in the Mott phase simply vanishes at the SI transition [see panels (a)-(d)]. On the other hand, as shown in Figs. 4(e) and 4(f), the excitation gap in the Mott phase becomes larger near the upper SI phase boundary [solid line in Fig. 1]. The gap remains in the rf-spectrum even at the SI phase transition. In the superfluid phase, in addition to the band with a finite threshold energy, a gapless spectrum appears in the rf-spectrum [see Figs. 4(g) and 4(h)]. This result directly comes from the dual structure of the hole excitations shown in Figs. 2(g) and 2(h). Namely, the rf-spectrum in the superfluid phase with $n\lesssim 1$ is very different from that with $n\gesim 1$. In particular, when $n\gesim 1$, we can observe the dual band structure of a strongly-correlated superfluid Bose gas by the rf-tunneling measurement.
\par
To summarize, we have discussed the excitation spectrum of a strongly-correlated Bose gas in an optical lattice. In the Mott phase, the Hubbard gap is obtained, reflecting the localized character of atoms. At the superfluid-Mott transition, particle or hole excitations become gapless while the other branch still has a finite energy gap. However, the latter branch also becomes gapless in the superfluid phase due to the appearance of a gapless band. The resulting spectrum has a dual structure, where itinerant gapless excitations and the localized band with a finite threshold energy coexist. We have also shown that this dual structure of the excitation spectrum may be observed by the rf-tunneling current measurement. Since the duality is absent in a weakly-interacting Bose gas, the observation of this phenomenon would be very useful in understanding the physics of a strongly-correlated superfluid Bose system. 
\par
This work was supported by a Grant-in-Aid for Scientific research from the Ministry of Education of Japan. 
%

%

\begin{thebibliography}{99}
\bibitem{note} In a $d$-dimensional lattice, we obtain $v_\phi/zt=\Phi\sqrt{2/d}$, which agrees with Ref.\cite{Krish} for a two-dimensional square lattice, giving $v_\phi/zt=\Phi$.
\end{thebibliography}
\end{document}